# Electronic Structure Modulation from Configuring Anatase $TiO_2$ into a Bicontinuous Mesostructure

Ying-Hao Lu, Bor Kae Chang,* and Yi-Fan Chen*

Department of Chemical and Materials Engineering, National Central University, Taoyuan 32001, Taiwan.

Email: yifanchen@ncu.edu.tw, BKChang@cc.ncu.edu.tw

**Abstract**

Configuring $TiO_2$ into bicontinuous mesostructures greatly improves its photocatalytic efficiency. This is often ascribed to the expanded surface area. Yet, whether mesostructuring modulates $TiO_2$'s electronic structure and how that contributes to the improvement are rarely discussed. Here, we employed spectroscopic and density functional theory approaches to address the question. It is found that the improved efficacy could arise from an expansion in surface area and elevation in density of states, both of which might collectively lead to the observed reduction in charge-carrier recombination.

Doping or structuring it into porous formats, with or without templates, is a common strategy to address issues arising from the relatively wide band gap (~3.2 eV) and high charge-carrier recombination rate of titanium oxide ($TiO_2$) [1,2], a popular metal-oxide semiconductor of low cost, nontoxicity, and high abundance [3-5]. The structuring approaches that involve templates enable a high degree of control over pore size, pore shape, pore spatial arrangement, and specific surface area $A_s$ for porous $TiO_2$; some approaches even produce porous $TiO_2$ featuring pores in the periodic mesostructures of bicontinous cubic phases [2]. The mesostructured $TiO_2$ was shown to exhibit considerably higher photocatalytic and quantum efficiencies, as catalysts and components of solar cells, respectively, than their nonporous counterparts [6-9]. The improvements are mostly ascribed to the $A_s$ expansion, due to the marked difference in $A_s$ between the mesostructured and nonporous $TiO_2$. However, whether the electronic structure is also modulated upon mesostructuring and involved in the mesostructuring-endowed improvements is rarely explored, even though such an involvement was noted for mesostructured $TiO_2$ deployed to dye-sensitized solar cells [10]. Understanding how mesostructuring modulates $TiO_2$'s electronic structure and thereby improves its performance may further allow one to tailor $TiO_2$ for diverse application scenarios. Herein, we employed a series of experimental and theoretical approaches to investigate how configuring anatase $TiO_2$ into the periodic mesostructure of a bicontinous cubic phase modulated the electronic structure and explored how the modulation might improve the photocatalytic efficacy.

Templated $TiO_2$ films were prepared with evaporation-induced self-assembly [11], where a template (here, the copolymer pluronic F127) self-assembled into a mesostructure in a solvent and a $TiO_2$ precursor polymerized to $TiO_2$ within the solvent channel of the mesostructure as the solvent evaporated [12]. F127 was subsequently removed from the composite films by calcination to obtain templated $TiO_2$ films. Nonporous $TiO_2$ films were prepared similarly but in the absence of F127. X-ray diffraction (XRD) confirms the anatase nature of $TiO_2$'s atomic lattice for both film types (Fig. S1). The mesoscale structure of the templated films was determined with small-angle X-ray scattering (SAXS). The scattering peaks of the films before and after calcination can be indexed to the space group of *Im3m* [Fig. 1(a)], consistent with the bicontinuous mesostructure, the primitive cubic phase, formed by F127 [13], verifying the formation of a bicontinuous mesostructure. The lattice parameter for the mesostructure of the films before and after calcination is, respectively, ~17.6 nm and ~13.7 nm [from the (110) reflection at $|\vec{Q}| \cong 0.5$ nm$^{-1}$ and 0.65 nm$^{-1}$, where $|\vec{Q}| = 4\pi \sin\theta / \lambda$, with $2\theta$ and $\lambda$ being the scattering angle and X-ray wavelength, respectively]. The reaction rate of the $TiO_2$-catalyzed methylene blue (MB) degradation was measured with UV absorption to evaluate the photocatalytic efficacy. Under the first-order reaction assumption, the measured MB concentration at a given time $t$ relative to the initial concentration $C(t)/C_0$ yielded the rate constants $k$ = 0.33 min$^{-1}$ and 0.09 min$^{-1}$ for mesostructured and nonporous $TiO_2$, respectively, a nearly quadruple improvement that may arise from the $A_s$ expansion and other factors [Fig. 1(b)].

The pore diameter and $A_s$ of mesostructured $TiO_2$ were then measured *via* $N_2$ sorption in accordance with the Brunauer−Emmett−Teller (BET) method [12]. The dominant pore diameter of mesostructured $TiO_2$ is ~6.3 nm [Fig. S2(b)], consistent with the mesoporous nature and the lattice parameter from SAXS [Fig. 1(a)]. The $A_s$ is 122.5 m$^2$ g$^{-1}$, about twice as large as that of nonporous $TiO_2$ (50 - 85 m$^2$ g$^{-1}$ [14,15]). The larger $A_s$ of mesostructured $TiO_2$ contributed to its higher photocatalytic efficacy. However, by only a factor of 2, the $A_s$ increase may not fully account for the 4-fold improvement in photocatalytic efficacy upon mesostructuring.

One additional factor that may have contributed to the quadruple improvement is changes in electronic structure. To explore this possibility, we examined the bang gap $E_g$ and the charge-carrier recombination for the $TiO_2$ films *via* UV absorption and photoluminescence (PL) spectroscopy, respectively [12]. The UV spectra of absorption coefficient $\alpha$ against the incident UV frequency $\nu$ was used to construct the Tauc plots of $(\alpha \cdot h\nu)^{1/\gamma}$ against $h\nu$, where $\gamma$ = 1/2 and 2 for direct and indirect band-gap semiconductors, respectively; and the *x*-intercept of the linear fit to the section showing a drastic increase in $(\alpha \cdot h\nu)^{1/\gamma}$ yielded $E_g$ [16]. For nonporous $TiO_2$, the obtained Tauc plot and $E_g$ (= 3.22 eV; Fig. S3) are in good agreement with those for anatase $TiO_2$, which possesses an indirect band gap, from the literature [16]. For mesostructured $TiO_2$, a careful analysis is warranted, given the uncertainty on whether mesostructuring transformed anatase $TiO_2$ to a direct band-gap semiconductor. Two Tauc plots, with $\gamma$ = 2 or 1/2, were thus constructed [Figs. 2(a) and 2(b)]. When treated as an indirect band-gap semiconductor, mesostructured $TiO_2$ shows a significant sub-band gap absorption on the Tauc plot [Fig. 2(a)]. The presence of this broad absorption band raises the possibility that intraband-gap states could have been introduced to $TiO_2$ [16-18], potentially by the defects specific to mesostructured $TiO_2$. Due to the sub-band gap absorption, $E_g$ cannot be determined simply from the *x*-intercept of the linear fit. Instead, additional linear fitting to the broad band is required so that $E_g$ can be derived from the intersection of the two linear fits [16]. $E_g$ thereby determined for mesostructured $TiO_2$ is 3.02 eV, marginally narrower than that of nonporous $TiO_2$. On the other hand, when treated as a direct band-gap semiconductor, mesostructured $TiO_2$ gave rise to a Tauc plot and $E_g$ [= 3.28 eV, Fig. 2(b)] essentially identical to those of nonporous $TiO_2$. Even though having no effect on $E_g$ in this scenario, mesostructuring still modulated the electronic structure and transform $TiO_2$ to a direct band-gap semiconductor. Altogether, the observations suggested that configuring $TiO_2$ into the bicontinuous mesostructure could perturb its electronic structure by introducing intraband-gap states or by turning it into a direct band-gap semiconductor. Nevertheless, both the modulations are unfavorable to photocatalysis (see below) [19,20]. Thus, they cannot account for the improved photocatalytic efficacy of mesostructured $TiO_2$. Other modulations in electronic structure might be present and underly the high efficacy, as revealed by PL.

PL spectra of both film types in Fig. 2(c) feature interference fringes characteristic of anatase $TiO_2$ films [20]. The spectra are also qualitatively similar, showing a broad peak centered at ca. 470 nm (or 2.64 eV) and a tail extending beyond 650 nm (or 1.91 eV), which, respectively, correspond to the green and red regions reported in earlier studies [20,21]. Given the spectral similarity, particularly in wavelengths where the peaks reside, it is likely that both nonporous and mesostructured $TiO_2$ are indirect band-gap semiconductors. However, the two spectra differ substantially in peak intensities: mesostructuring greatly reduced the intensity by approximately a factor of 2 at the spectral maximum of ca. 470 nm [Fig. 2(c)]. The reduction suggests a lower radiative recombination rate or longer-lasting electron-hole separation for mesostructured $TiO_2$ than for nonporous $TiO_2$. Given that photogenerated holes in the valence band (VB) play a key role in $TiO_2$-photocatalyzed reactions [22], the longer-lasting electron-hole separation is expected to be crucial for the greater photocatalytic efficiency of mesostructured $TiO_2$. Hence, digging into the mechanisms underlying the PL intensity reduction should shed light on how mesostructuring led to the high photocatalytic efficacy, in addition to expanding $A_s$.

As an indirect band-gap semiconductor, anatase $TiO_2$ cannot radiate efficiently *via* direct recombination of electrons from the conduction band (CB) and holes from VB. Instead, PL of anatase $TiO_2$ relies on other mechanisms to attain radiative recombination, including those involving polarons, *i.e.*, charge carriers trapped in the intraband-gap states arising from surface or subsurface defects [20,21,23]. Based on the findings of Palloti et al. [20], we may ascribe the broad peak centered at ca. 470 nm in Fig. 2(c) to radiative recombination of photogenerated electrons in CB with holes trapped in oxygen vacancies at the surface. On the other hand, radiative recombination of holes in VB with electrons trapped in the intraband-gap states from subsurface oxygen vacancies may account for the tail extending beyond 650 nm. With this understanding, three potential mechanisms are considered to underly the PL intensity reduction for mesostructured $TiO_2$. One involves the expanded $A_s$ afforded by mesostructuring. Since $O_2$ adsorbed on $TiO_2$ surface captures electrons in CB and prevents them from recombining with holes in the trap states [20,21,24], an $A_s$ expansion can suppress PL as it provides more sites for $O_2$ adsorption. Accordingly, the expanded $A_s$ by mesostructuring offered not only larger estate for the photocatalytic reaction but also more adsorbed $O_2$ to sequester electrons in CB, resulting in longer-lasting holes in VB. Another mechanism concerns the suppression of surface and subsurface defects, which reduces the amount of the trapped holes and electrons and thus depresses PL. However, mesostructuring is considered to spawn more defects, owing to potential lattice distortion by mesostructuring; and the prospect of reducing surface defects and thus the intraband-gap states by mesostructuring is also inconsistent with the presence of the broad absorption band on the Tauc plot in Fig. 2(a). Hence, the other viable mechanism of the three is the mesostructuring-mediated modulation in electronic structure which alters the electron population distribution and the availability of energy states in VB and CB. The modulation could possibly better sustain free charge carriers or provide charge carriers more ways to escape the trap states.

To learn how mesostructuring might attain the electronic structure modulation near the band gap, we carried out density functional theory (DFT) calculations to obtain the density of states (DOS) for nonporous and mesostructured $TiO_2$ [12]; with the electron population being the product of DOS and Fermi-Dirac statistics, acquiring DOS would elucidate how mesostructuring might have shifted the population distribution. As shown in Figs. 3 and S4, the DOS for nonporous $TiO_2$ demonstrates features known for anatase $TiO_2$ [23]: The band gap between VB maximum (VBM) and CB minimum (CBM) is 3.39 eV (Fig. 3), in good agreement with $E_g$ reported in the literature [16] and derived from the Tauc plot here (Fig. S3); and CB and VB are, respectively, dominated by Ti 3d states and O 2p states. After mesostructuring, $E_g$ of $TiO_2$ (=3.50 eV) was essentially unchanged (Fig. 3), consistent with the finding from the Tauc plots [Figs. 2(a), 2(b), and S3]. Nevertheless, nonporous and mesostructured $TiO_2$ differ considerably in the DOS near VBM and CBM, with mesostructured $TiO_2$ possessing a significantly higher DOS (Fig. 3). Increases in O 2p DOS in both VBM and CBM, and Ti 3d DOS increases in CBM are mostly responsible for the overall DOS elevation. As a result, in mesostructured $TiO_2$, O 2p states remain dominant in VB, but both Ti 3d and O 2p characteristics are found in CB (Fig. S4). Interestingly, DOS for different spin orientations differ slightly near CBM, signifying ferromagnetism for mesostructured $TiO_2$. However, examining the magnetism is out of the scope of the present study.

It has to be noted that mesostructuring *per se* is not believed to have introduced defects in $TiO_2$'s atomic lattice and produced the intraband-gap states evidenced in Fig. 2(a). This is because, if existing, the mesostructuring-induced intraband-gap states would have been revealed by the calculations and manifested in the DOS. Their absence in the DOS suggests the experimentally observed intraband-gap states to have arisen from the defects introduced by one or more of the sample preparation processes (*e.g.*, calcination), rather than by configuring the atomic lattice into a mesostructure.

As revealed by the calculations, the mesostructuring-mediated DOS elevation could increase the electron population and the accessible energy states near VBM and CBM. Through the two effects, the DOS elevation of mesostructured $TiO_2$ might lead to the observed reduction in charge-carrier recombination [Fig. 2(c)], a possibility supported by earlier studies [25,26], and contribute to the observed improvement in photocatalytic efficacy [Fig 1(b)]. This might be realized on two fronts. First, the larger electron population near VBM supplied more electrons to populate the energy states in CB upon photoexcitation and might thus lead to more photogenerated electrons in CB and correspondingly more holes in VB. Given the importance of holes in $TiO_2$-photocatalyzed reactions [22], the increased hole population in VB should benefit the photocatalytic reaction. Second, it is speculated that more energy states near the band gap provided holes and electrons trapped in the intraband-gap states (*i.e.*, polarons), which were spawned from defects possibly introduced by sample processing, more ways to escape or be excited to VB and CB, respectively. Since the radiative recombination involves the polarons (see above) [20,21,23], escape of the trapped charge carriers might close the routes whereby the charge carriers radiatively recombine with their free counterparts and allow holes in VB to take part in the photocatalytic reaction. The two fronts might synergistically facilitate the photocatalytic reaction for mesostructured $TiO_2$.

In summary, we employed the copolymer pluronic F127 as the template to configure anatase $TiO_2$ into a bicontinuous mesostructure with the space group of *Im3m* and confirmed the structural features with SAXS and XRD. Compared to its nonporous counterpart, mesostructured $TiO_2$ displayed a quadruply higher photocatalytic efficacy in degrading methylene blue. Through BET sorption, UV absorption and PL spectroscopy, and DFT calculations, we found that mesostructuring altered the electronic structure of anatase $TiO_2$ and argue that the 4-fold improvement arose not only from the doubly expanded $A_s$ but also from the mesostructuring-induced DOS elevation, which might suppress radiative recombination of charge carriers. It is likely that these factors acted in unison or even synergistically: In addition to providing more sites for the reaction, the larger $A_s$ might also lead to more adsorbed $O_2$ on the $TiO_2$ surface, which could better sequester the increased electron population in CB due to the DOS elevation and thereby allow the correspondingly increased hole population in VB to last longer for the photocatalysis. Though sample processing for mesostructured $TiO_2$ appears to introduce defects and thus intraband-gap states, which would have facilitated the radiative recombination and thus obstructed the photocatalysis, the collective effects of the expanded $A_s$ and the elevated DOS apparently overrode that of the intraband-gap states. The findings may apply to other metal-oxide semiconductors and highlight the importance of taking into consideration both modified $A_s$ and altered electronic structure to fully understand and predict how mesostructuring changes the performance of these semiconductors.

**Acknowledgement**

This study is supported by the National Science and Technology Council (Grant No. MOST-111-2221-E-008-003-).

**Reference**

1. W. Zhang, H. He, H. Li, L. Duan, L. Zu, Y. Zhai, W. Li, L. Wang, H. Fu, and D. Zhao, Visible-Light Responsive TiO2-Based Materials for Efficient Solar Energy Utilization, Adv. Energy Mater. **11,** 2003303 (2020).
2. D. Fattakhova-Rohlfing, A. Zaleska, and T. Bein, Three-Dimensional Titanium Dioxide Nanomaterials, Chem. Rev. **114,** 9487 (2014).
3. D. Chen, Y. Cheng, N. Zhou, P. Chen, Y. Wang, K. Li, S. Huo, P. Cheng, P. Peng, R. Zhang, L. Wang, H. Liu, Y. Liu, and R. Ruan, Photocatalytic degradation of organic pollutants using TiO2-based photocatalysts: A review, J. Clean. Prod. **268,** 121725 (2020).
4. M. Kokkonen, P. Talebi, J. Zhou, S. Asgari, S. A. Soomro, F. Elsehrawy, J. Halme, S. Ahmad, A. Hagfeldt, and S. G. Hashmi, Advanced research trends in dye-sensitized solar cells, J. Mater. Chem. A **9,** 10527 (2021).

5. Q. Ni, R. Dong, Y. Bai, Z. Wang, H. Ren, S. Sean, F. Wu, H. Xu, and C. Wu, Superior sodium-storage behavior of flexible anatase TiO2 promoted by oxygen vacancies, Energy Storage Mater. **25,** 903 (2020).

6. H. Choi, E. Stathatos, and D. D. Dionysiou, Synthesis of nanocrystalline photocatalytic TiO2 thin films and particles using sol–gel method modified with nonionic surfactants, Thin Solid Films **510,** 107 (2006).

7. J. Wang, H. Li, H. Li, C. Zou, H. Wang, and D. Li, Mesoporous $TiO_2$ thin films exhibiting enhanced thermal stability and controllable pore size: preparation and photocatalyzed destruction of cationic dyes, ACS Appl. Mater. Interfaces **6,** 1623 (2014).

8. Y. Xiong, D. He, R. Jaber, P. J. Cameron, and K. J. Edler, Sulfur-Doped Cubic Mesostructured Titania Films for Use as a Solar Photocatalyst, J. Phys. Chem. C **121,** 9929 (2017).

9. M. Nedelcu, S. Guldin, M. C. Orilall, J. Lee, S. Hüttner, E. J. W. Crossland, S. C. Warren, C. Ducati, P. R. Laity, D. Eder, U. Wiesner, U. Steiner, and H. J. Snaith, Monolithic route to efficient dye-sensitized solar cells employing diblock copolymers for mesoporous TiO2, J. Mater. Chem. **20,** 1261 (2010).

10. P. Docampo, S. Guldin, M. Stefik, P. Tiwana, M. C. Orilall, S. Hüttner, H. Sai, U. Wiesner, U. Steiner, H. J. Snaith, Control of Solid-State Dye-Sensitized Solar Cell Performance by Block-Copolymer-Directed TiO2 Synthesis, Adv. Funct. Mater. **20,** 1787 (2010).

11. T. S. Dörr, L. Deilmann, G. Haselmann, A. Cherevan, P. Zhang, P. Blaha, P. W. de Oliveira, T. Kraus, D. Eder, Ordered Mesoporous TiO2 Gyroids: Effects of Pore Architecture and Nb-Doping on Photocatalytic Hydrogen Evolution under UV and Visible Irradiation, Adv. Energy Mater. **8,** 1802566 (2018).

12. See Supplemental Material for technical details on the experiments and calculations and for additional data characterizing the $TiO_2$ films.

13. D. Zhao, Q. Huo, J. Feng, B. F. Chmelka, and G. D. Stucky, Nonionic Triblock and Star Diblock Copolymer and Oligomeric Surfactant Syntheses of Highly Ordered, Hydrothermally Stable, Mesoporous Silica Structures, J. Am. Chem. Soc. **120,** 6024 (1998).

14. J. Y. Park, H.-H. Kim, D. Rana, D. Jamwal, and A. Katoch, Surface-area-controlled synthesis of porous TiO2 thin films for gas-sensing applications, Nanotechnology **28,** 095502 (2017).

15. K. Maver, U. L. Štangar, U. Černigoj, S. Gross, and R. C. Korošec, Low-temperature synthesis and characterization of TiO2 and TiO2–ZrO2 photocatalytically active thin films, Photochem. Photobiol. Sci. **8,** 657 (2009).

16. P. Makuła, M. Pacia, and W. Macyk, How To Correctly Determine the Band Gap Energy of Modified Semiconductor Photocatalysts Based on UV–Vis Spectra, J. Phys. Chem. Lett. **9,** 6814 (2018).

17. G. Sarigul, I. Chamorro-Mena, N. Linares, J. García-Martínez, E. Serrano, Hybrid Amino Acid-$TiO_2$ Materials with Tuneable Crystalline Structure and Morphology for Photocatalytic Applications, Adv. Sustainable Syst. **5,** 2100076 (2021).

18. H. Ali-Löytty, M. Hannula, J. Saari, L. Palmolahti, B. D. Bhuskute, R. Ulkuniemi, T. Nyyssönen, K. Lahtonen, and M. Valden, Diversity of $TiO_2$: Controlling the Molecular and Electronic Structure of Atomic-Layer-Deposited Black $TiO_2$, ACS Appl. Mater. Interfaces **11,** 2758 (2019).

19. J. Zhang, P. Zhou, J. Liu, and J. Yu, New understanding of the difference of photocatalytic activity among anatase, rutile and brookite $TiO_2$, Phys. Chem. Chem. Phys. **16,** 20382 (2014).

20. D. K. Pallotti, L. Passoni, P. Maddalena, F. Di Fonzo, and S. Lettieri, Photoluminescence Mechanisms in Anatase and Rutile $TiO_2$, J. Phys. Chem. C **121,** 9011 (2017).

21. C. Mercado, Z. Seeley, A. Bandyopadhyay, S. Bose, and J. L. McHale, Photoluminescence of Dense Nanocrystalline Titanium Dioxide Thin Films: Effect of Doping and Thickness and Relation to Gas Sensing, ACS Appl. Mater. Interfaces **3,** 2281 (2011).

22. Y. Kakuma, A. Y. Nosaka, and Y. Nosaka, Difference in $TiO_2$ photocatalytic mechanism between rutile and anatase studied by the detection of active oxygen and surface species in water, Phys. Chem. Chem. Phys. **17,** 18691 (2015).

23. J. J. Carey and K. P. McKenna, Does Polaronic Self-Trapping Occur at Anatase $TiO_2$ Surfaces? J. Phys. Chem. C **122,** 27540 (2018).

24. R. Brüninghoff, K. Wenderich, J. P. Korterik, B. T. Mei, G. Mul, and A. Huijser, J. Phys. Chem. C **123,** 26653 (2019).

25. S. K. Pathak, A. Abate, P. Ruckdeschel, B. Roose, K. C. Gödel, Y. Vaynzof, A. Santhala, S.-I. Watanabe, D. J. Hollman, N. Noel, A. Sepe, U. Wiesner, R. Friend, H. J. Snaith, and U. Steiner, Performance and Stability Enhancement of Dye-Sensitized and Perovskite Solar Cells by Al Doping of $TiO_2$, Adv. Funct. Mater. **24,** 6046 (2014).

26. C. Regmi, Y. K. Kshetri, T.-H. Kim, D. Dhakal, S. W. Lee, Mechanistic understanding of enhanced photocatalytic activity of N-doped $BiVO_4$ towards degradation of ibuprofen: An experimental and theoretical approach, Mol. Catal. **470,** 8 (2019).

**Fig. 1.** Mesoscale-structural (a) and catalytic (b) characterizations of the $TiO_2$ films. Miller indices in (a) indicate the expected positions of the scattering peaks for the space group *Im3m*. Lines in (b) are fits of $-\ln(C/C_0) = kt$ for the first-order reaction, where $C/C_0$ is the residual MB concentration relative to the initial concentration, $k$ is the rate constant, and $t$ is the aggregate reaction time.

**Fig. 2.** Tauc plots (a,b) and PL spectra (c) of the $TiO_2$ films. Mesostructured $TiO_2$ is treated as an indirect (a) or direct (b) band-gap semiconductor.

**Fig. 3.** Total DOS for nonporous and mesostructured $TiO_2$.

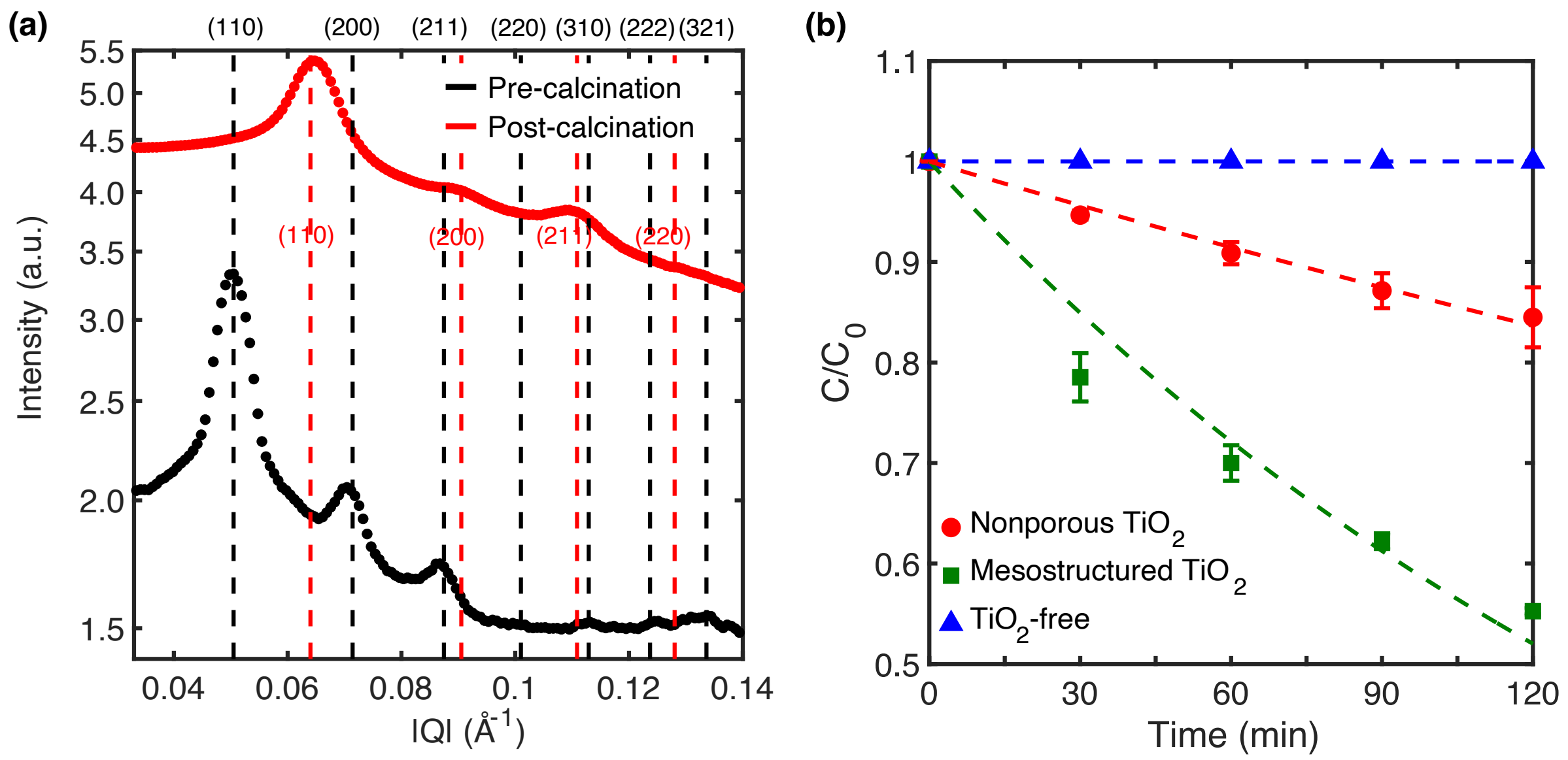

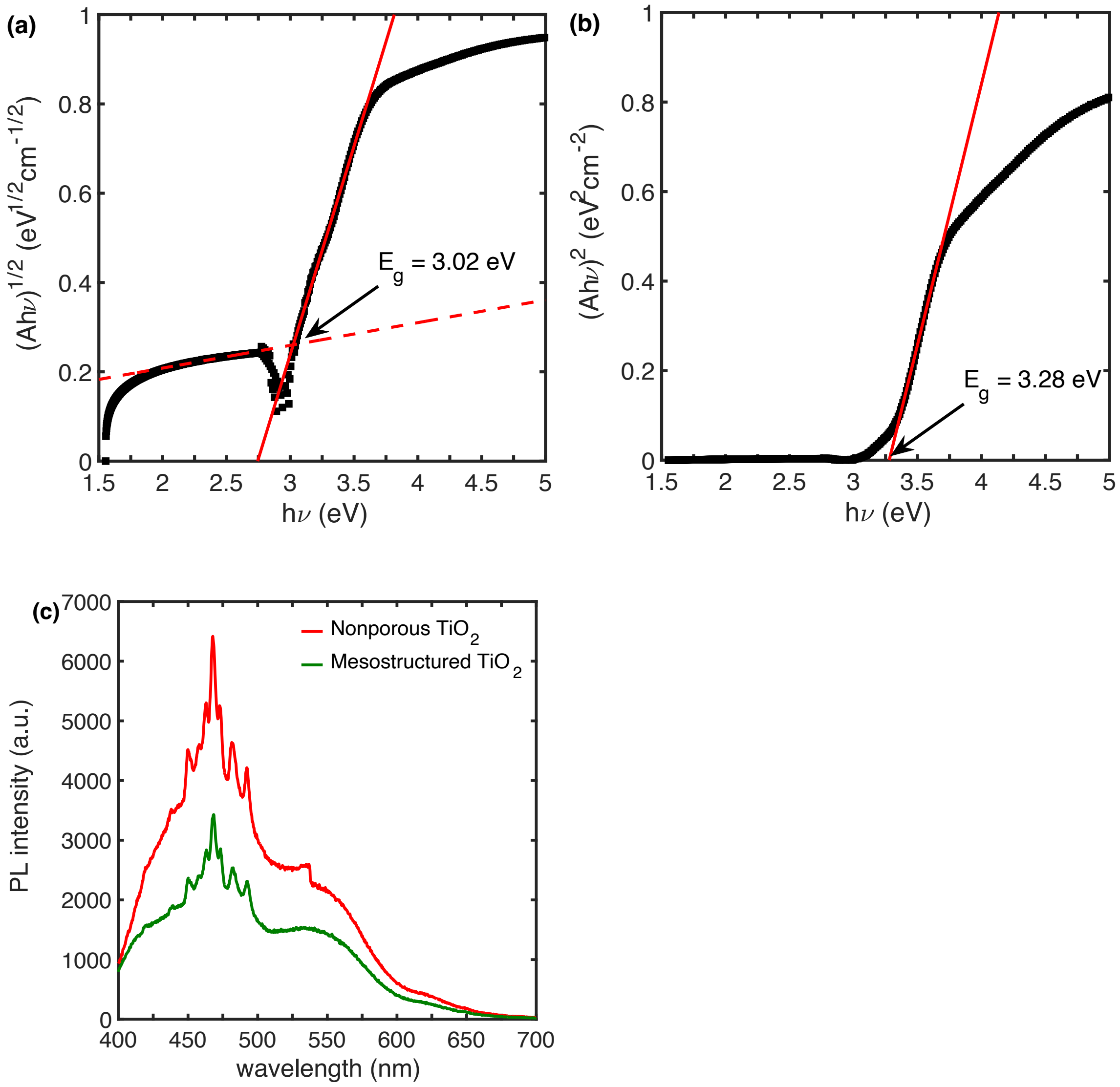

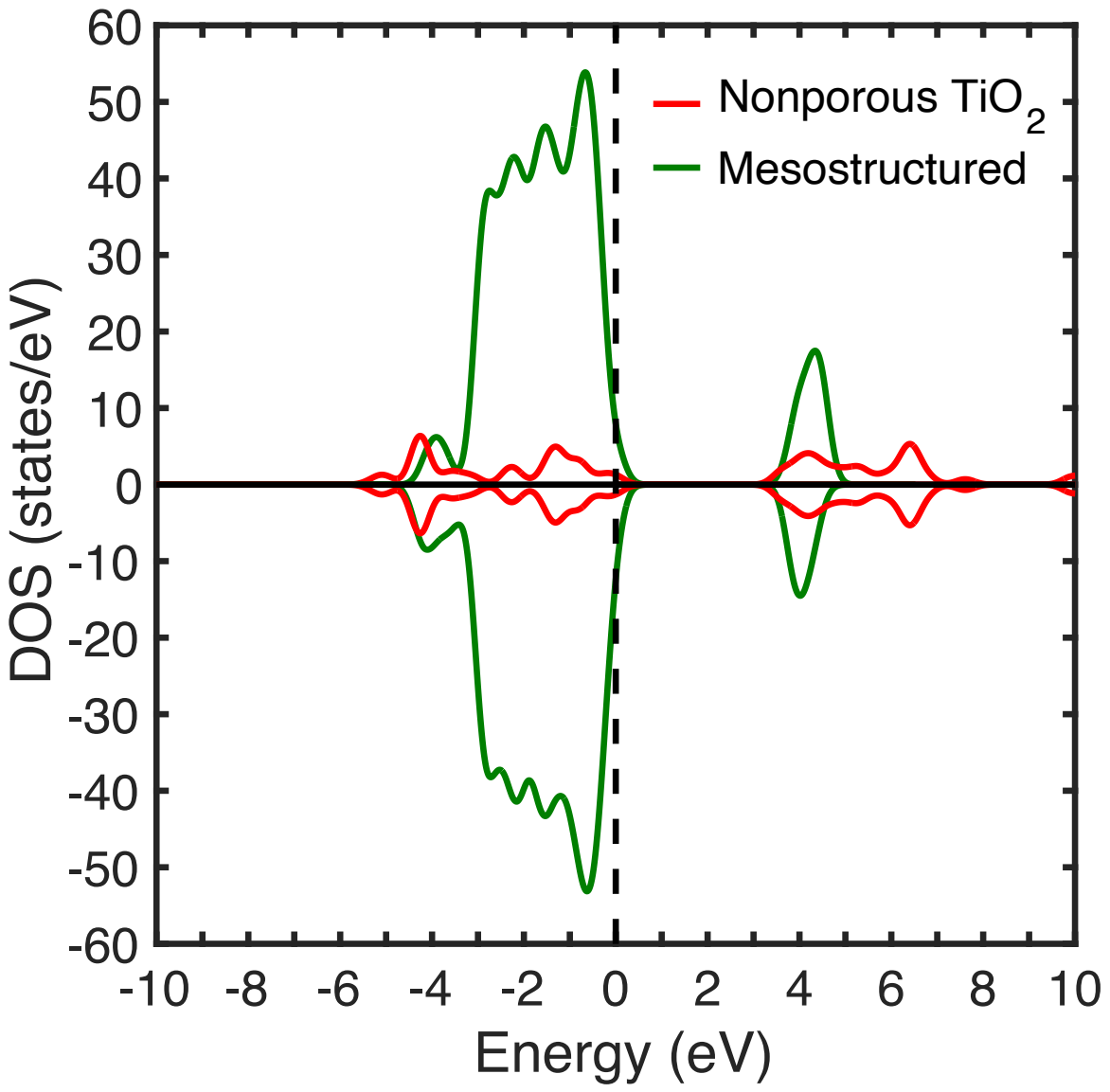

# Supplemental Material: Electronic Structure Modulation from Configuring Anatase $TiO_2$ into a Bicontinuous Mesostructure

Ying-Hao Lu, Bor Kae Chang,* and Yi-Fan Chen*

Department of Chemical and Materials Engineering, National Central University, Taoyuan 32001, Taiwan.

**Materials**

The $TiO_2$ precursor, titanium(IV) isopropoxide (TTIP; >98%, cat. no. AC194700050), and the polymer template, Pluronic F127 (99.9%, cat. no. P2443), were sourced from Acros Organics and Sigma-Aldrich, respectively, and used as received. Other chemicals, including ethanol (HPLC grade, cat. no. 10538071, J.T. Baker) and hydrochloric acid (32%, cat. no. 30720, Sigma-Aldrich), were also used without further treatments.

**Preparation of mesostructured $TiO_2$ films**

Mesostructured $TiO_2$ films were prepared with evaporation-induced self-assembly, following a well-established protocol [S1]. Briefly, F127 was dissolved in ethanol at 35°C. Deionized water and hydrochloric acid were successively mixed with the F127 solution before TTIP was added dropwise at 35°C. The constituents of the solution were finally in the molar ratio of TTIP/F127/ water/ethanol/HCl = 1/0.003/15/30/1.5. The solution was then placed in a peri dish and incubated at 35°C and 75.5% RH for 48 hours. During incubation, the solvent evaporated and F127 self-assembled into a mesostructure while TTIP polymerized to form amorphous $TiO_2$ within the solvent channel of the mesostructure. A dried film formed on the bottom of the peri dish and was subsequently calcinated by heating the film to 120°C at 1.5°C/min under Ar gas and to 350°C at 0.5°C/min under air to remove F127. The film was maintained at 350°C for 4 hours and then cooled to room temperature at 1.5°C/min; a $TiO_2$ film featuring a mesostructure was thereby acquired. Nonporous $TiO_2$ films were prepared similarly, except that F127 was not added to the solution.

**Characterization of the $TiO_2$ film**

The $TiO_2$ films were structurally characterized with small-angle X-ray scattering (SAXS), X-ray diffraction (XRD), and $N_2$ sorption for their mesostructure, atomic lattice, and specific surface area, respectively. SAXS data were collected from the $TiO_2$ films peeled from the peri dishes in the transmission geometry at Beamlines TLS-13A (wavelength = 1.02 Å; Rayonix SX165 MarCCD detector), TLS-23A (wavelength = 0.83 Å; Dectris Pilatus 1M-F pixel detector), and TPS-13A (wavelength = 0.83 Å; Dectris Eiger X 9M pixel detector) of National Synchrotron Radiation Research Center in Hsinchu, Taiwan or with a Cu$K_\alpha$ radiation from a rotating anode (wavelength = 1.54 Å; Rigaku Nano Viewer). The collected 2-D SAXS images were azimuthally integrated and corrected for background scattering and X-ray absorption to obtain scattering profiles of scattering intensity $I$ against momentum transfer $|\vec{Q}| = 4\pi \sin\theta / \lambda$, where $2\theta$ is the scattering angle and $\lambda$ is the wavelength of the incident X-ray. XRD was carried out with a diffractometer (Cu$K_\alpha$ radiation; Bruker D8 Advance), and the $N_2$ sorption isotherms were obtained with a gas sorption analyzer (Micromeritics TriStar 3000) to determine the specific surface areas in accordance with the Brunauer−Emmett−Teller (BET) method.

The electronic structure of the $TiO_2$ films was characterized by measuring the UV absorption and photoluminescence (PL) of the films to yield the band gap and recombination rate of the electron-hole pairs, respectively. The former was carried out with a UV-visible spectrometer (model no. UV-1900, Shimadzu). Tauc plots of $(\alpha \cdot h\nu)^{1/\gamma}$ against $h\nu$, where $\gamma$ = 1/2 and 2 for direct and indirect band gaps ($\gamma$ = 1/2 for $TiO_2$ [S2]), respectively, were drawn based on the measured absorption coefficient $\alpha$ and the incident UV frequency $\nu$. A linear function was fitted to the linear section of the plot, from which the band gap $E_g$ was estimated according to $(\alpha \cdot h\nu)^{1/\gamma} = B\left(h\nu - E_g\right)$, with $B$ being a fitting constant [S2]. PL of the $TiO_2$ films was excited by a laser with a wavelength of 325 nm and collected at wavelengths from 400 to 600 nm with a fluorometer (model no. FP-8300, JASCO).

**Photocatalysis**

The photocatalytic efficiency of the $TiO_2$ films was assessed by measuring the rate of the $TiO_2$-catalyzed methylene blue (MB) degradation. A $TiO_2$ film weighing 30 mg was placed into 10 ml of 30 $\mu$M MB solution. The sample was stored in the dark for 24 hours and subsequently subjected to UV irradiation at 254 nm to initiate the photodegradation. The MB concentration change as a function of time was recorded with a UV-visible spectrometer (model no. UV-1900, Shimadzu) *via* measuring the UV absorption at 664 nm every 30 minutes. With the relation based on the Beer-Lambert law and the first-order reaction assumption, $-\ln(A/A_0) = -\ln(C/C_0) = kt$, where $C$ and $A$ are the MB concentration and UV absorbance at a given time, $C_0$ and $A_0$ are the initial MB concentration and UV absorbance, respectively, $k$ is the first-order rate constant, and $t$ is the aggregate irradiation time, the MB concentration change and photodegradation rate were obtained.

**Theoretical calculations**

Calculations were carried out using the CASTEP software package [S3] for density functional theory (DFT) simulations utilizing plane-wave basis sets. Ultrasoft pseudopotentials were used at the GGA level of theory with the Perdew-Burke-Ernzerhof (PBE) density functional [S4]. A cutoff energy of 820 eV and gamma point were chosen, and convergence criteria of $10^{-5}$ eV/atom, 0.03 eV/Å, and 0.001 Å were designated for the total energy per atom, maximum ionic force, and maximum ionic displacement, respectively. A U value of 9.0 eV for Ti and 2×2×2 k-point grid were used for GGA+U electronic structure calculations of density of states (DOS) and partial density of states (PDOS).

**Reference**

S1. T. S. Dörr, et al., *Adv. Energy Mater*. **2018,** *8*, 1802566.

S2. P. Makuła, et al., *J. Phys. Chem. Lett.* **2018,** *9*, 6814.

S3. S. J. Clark, et al., *Zeitschrift für Kristallographie - Crystalline Materials* **2005,** *9*, 567.

S4. J. P. Perdew, et al., *Phys. Rev. Lett.* **1996,** *77*, 3865.

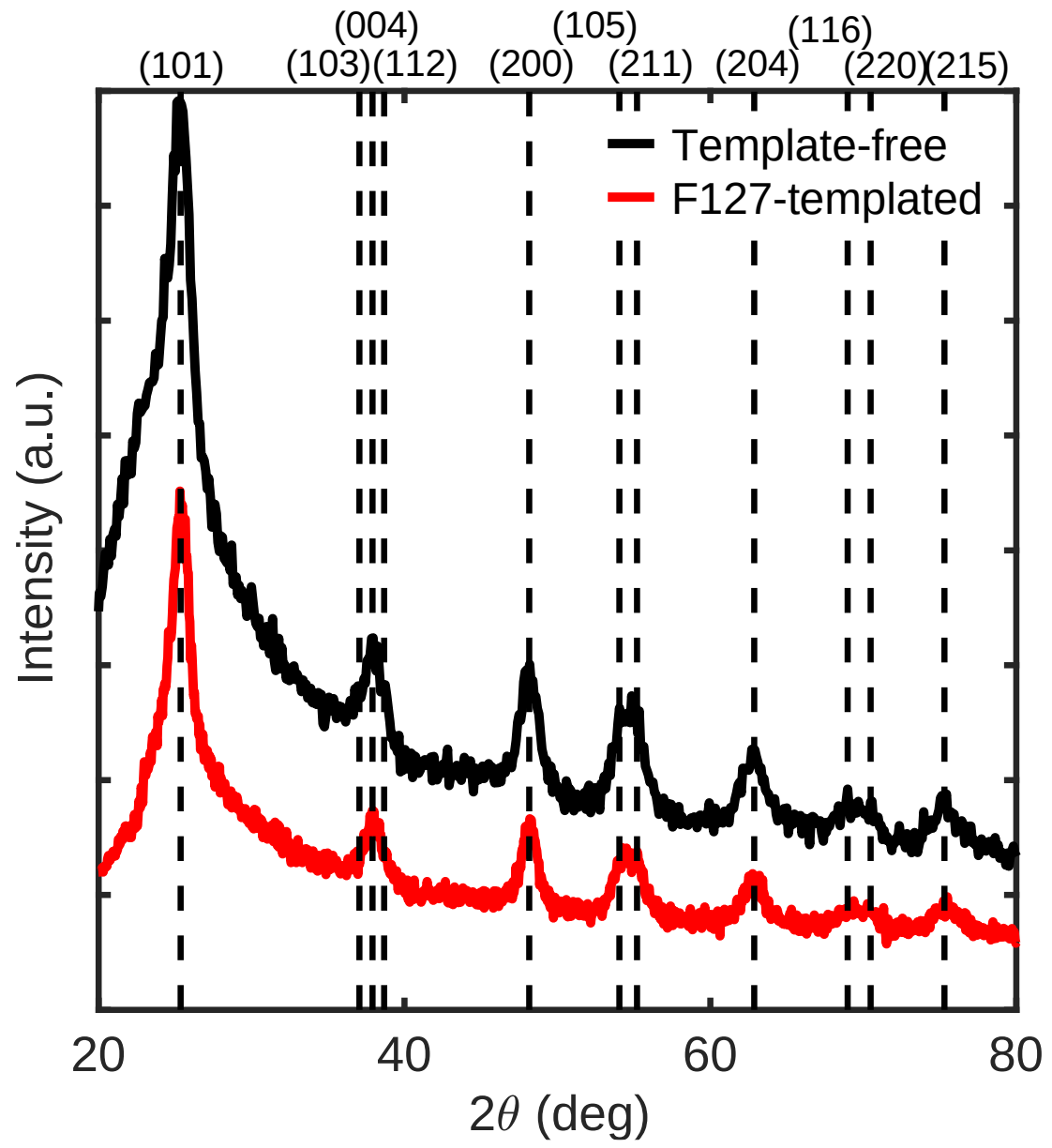

**Fig. S1.** XRD profile for the $TiO_2$ films with and without templating. Dashed lines mark the expected positions of the diffraction peaks for the anatase phase of $TiO_2$.

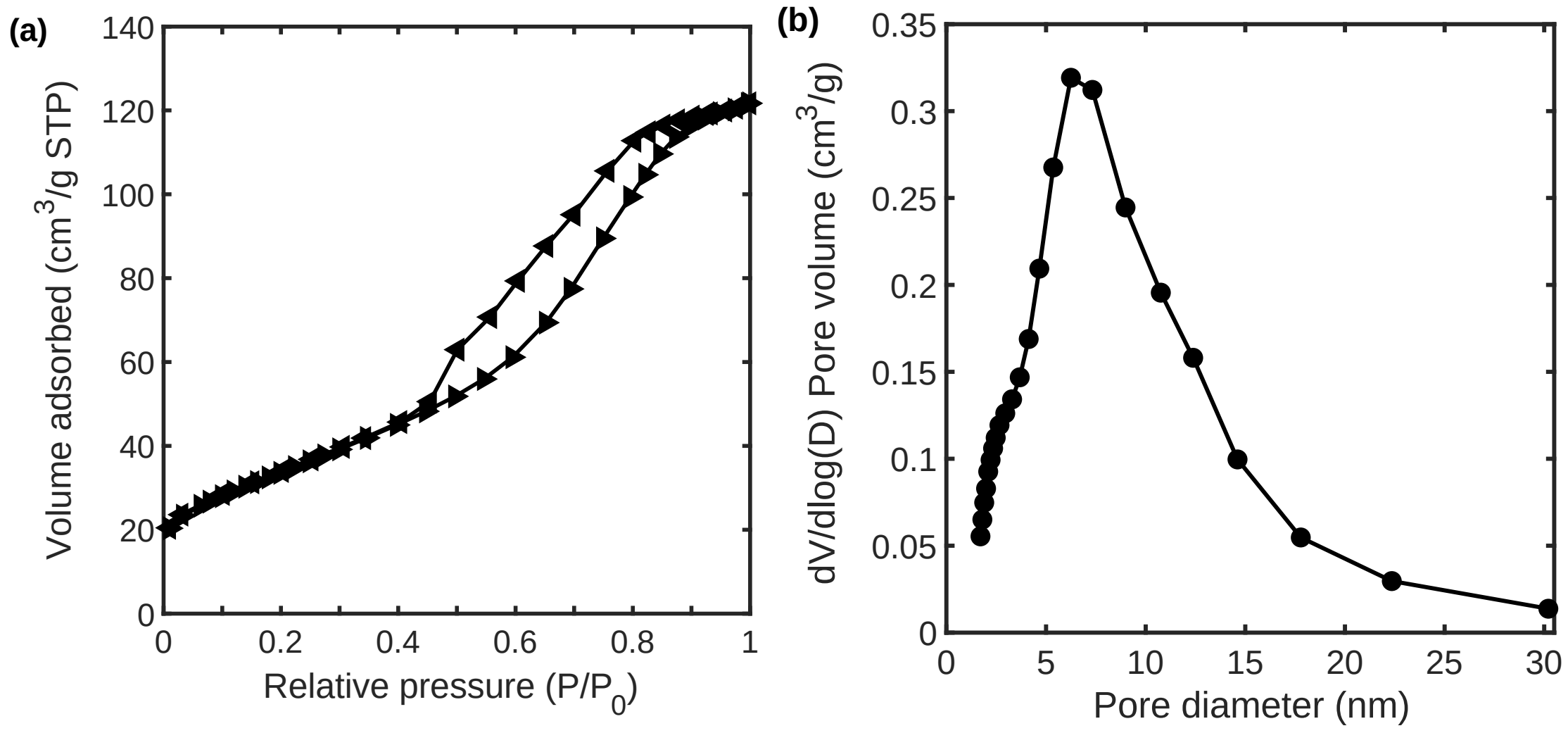

**Fig. S2.** $N_2$ sorption isotherms (a) and pore size distribution (b) of mesostructured $TiO_2$.

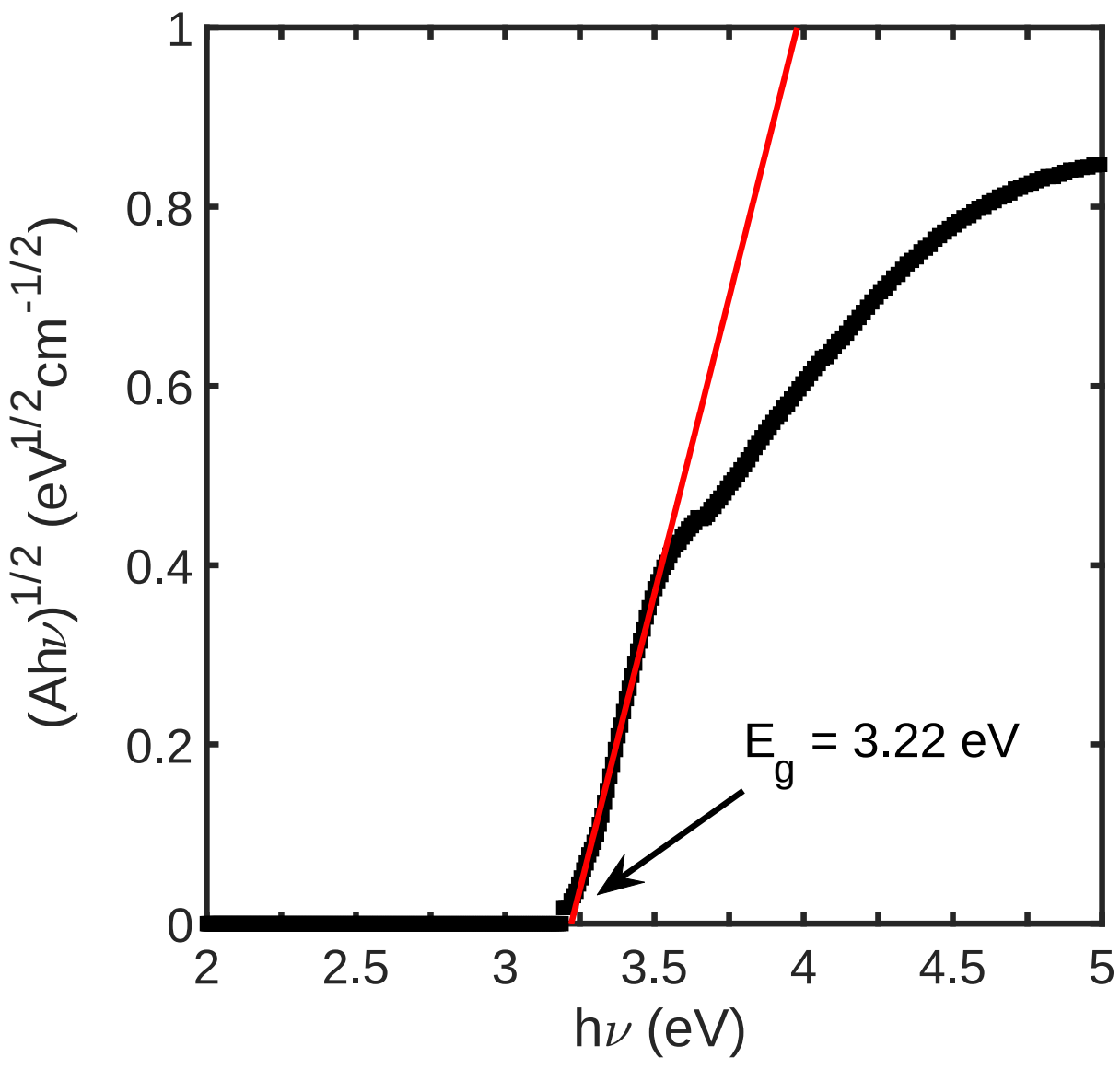

**Fig. S3.** Tauc plot for nonporous $TiO_2$. The *x*-intercept of the linear fit yields the band gap.

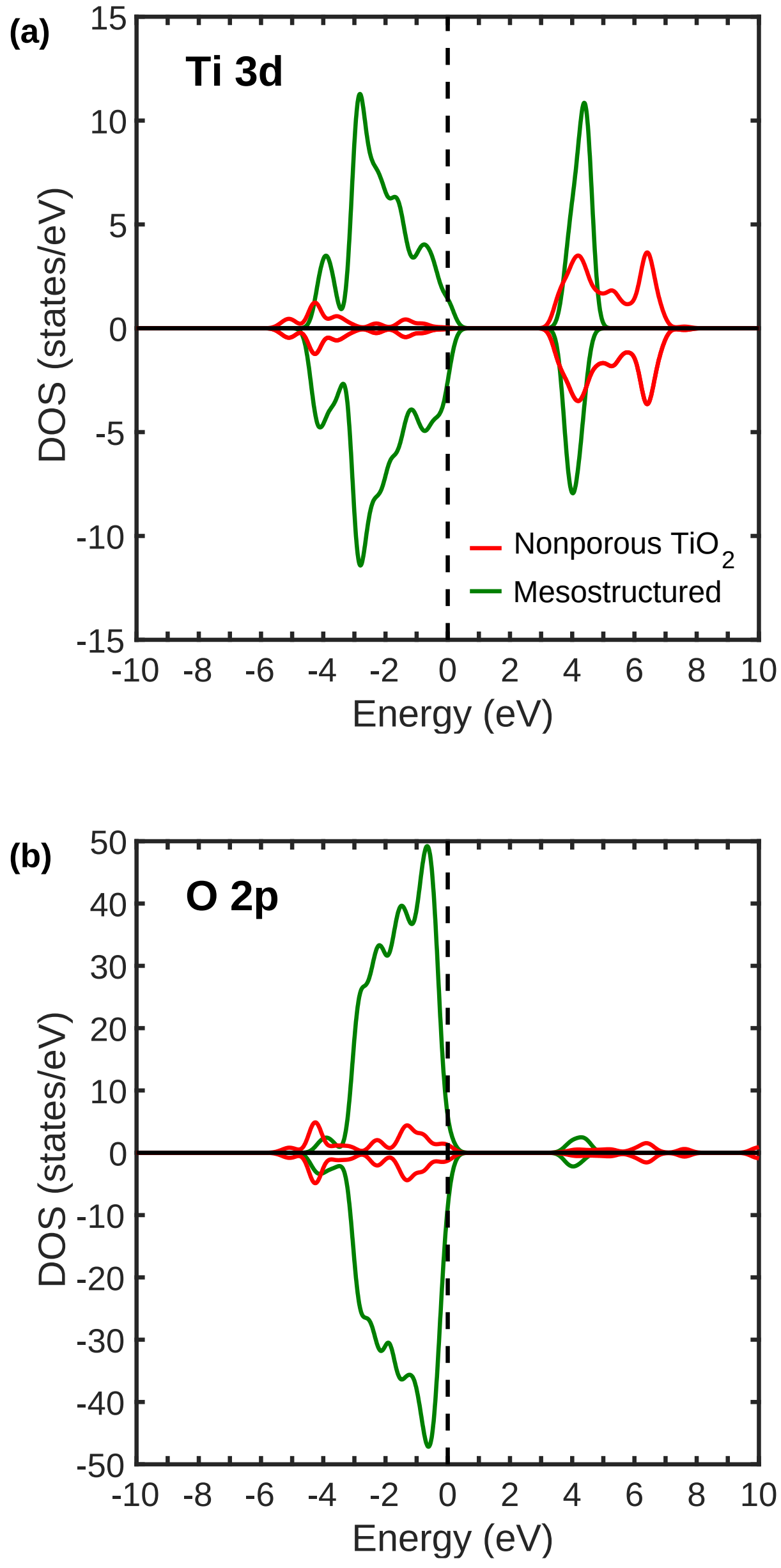

**Fig. S4.** Ti 3d (a) and O 2p (b) partial DOS obtained from the DFT calculations for nonporous and mesostructured $TiO_2$.